\documentclass[pra,twocolumn,showpacs]{revtex4}
\usepackage{graphicx}
\usepackage{amsmath}
\def\Journal#1#2#3#4{{#1} {\bf #2}, #3 (#4)}
\def\PR{Phys. Rev.}

\def\PRA{Phys. Rev. A}

\def\JMO{J. Mod. Opt.}
\newcommand{\n}{\nonumber}
\newcommand{\bn}{\begin{eqnarray}}
\newcommand{\en}{\end{eqnarray}}
\newcommand{\eml}{\end{multline}}
\newcommand{\bml}{\begin{multline}}
\newcommand{\h}{\hspace}

\begin{document}

\title {A Langevin analysis of fundamental noise limits in Coherent Anti-Stokes Raman Spectroscopy}
 \author{Kunal K. Das$^1$\footnote{Present Address:  Department of Physics, The
Pennsylvania State University, University Park, Pennsylvania
16802.\\
Send all correspondence to: kdas@phys.psu.edu}, G.S. Agarwal$^2$,
Yu. M. Golubev$^3$, and M.O. Scully$^{4,5}$}
 \affiliation{$^1$Institute for Quantum Studies and Department of Physics, Texas A$\&$M University, College Station, Texas 77843}
 \affiliation{$^2$Physical Research Laboratory, Navrangpura, Ahmedabad-380 009, India}
 \affiliation{$^3$V. A. Fock Physics Institute, St. Petersburg State
 University,ul. Ul'anovskaya 1,
 198504 St. Petersburg, Stary Petershof, Russia}
 \affiliation{$^4$Institute for Quantum Studies and Departments of Physics and Electrical and Chemical Engineering, Texas A$\&$M University, College Station,
Texas 77843}
 \affiliation{$^5$Departments of Chemistry and Aerospace and Mechanical Engineering, Princeton University, Princeton, NJ 08544}

\date{\today }
\begin{abstract}

We use a Langevin approach to analyze the quantum noise in
Coherent Anti-Stokes Raman Spectroscopy (CARS) in several
experimental scenarios: with continuous wave input fields acting
simultaneously and with fast sequential pulsed lasers where one
field scatters off the coherence generated by other fields; and
for interactions within a cavity and in free space.  In all the
cases, the signal as well as the quantum noise due to spontaneous
decay and decoherence in the medium are shown to be described by
the same general expression. Our theory in particular shows that
for short interaction times, the medium noise is not important and
the efficiency is limited only by the intrinsic quantum nature of
the photon. We obtain fully analytic results \emph{without} making
an adiabatic approximation, the fluctuations of the medium and the
fields are self solved consistently.
\end{abstract}
 \maketitle

\section{Introduction}

The field of Raman spectroscopy is a very mature one that boasts a
vast literature which over time has developed and explored
countless ingenious improvements and techniques to tweak out
better and stronger signals. Much of the motivation for this lies
in the fact that Raman scattering inevitably involves the
structure of the scattering medium, because the incoming and
outgoing signals differ by the frequency of some internal mode of
the constituent molecules or atoms, thereby making it an
invaluable tool for spectroscopy.

The weak signals associated with spontaneous Raman scattering were
long overcome by stimulated scattering through non-linear
interactions. Signal has been enhanced through resonance with some
specific natural modes of the molecules. In particular the
technique of Coherent Anti-Stokes Raman Scattering (CARS) has
emerged as one of the most useful Raman technique; it involves two
incoming fields that create coherence in the medium which thereby
enhances the scattered signal. The generated field being
anti-Stokes eliminates fluorescence problems, and  resonance
enhancement is also usually possible.

One way to improve on existing CARS techniques has been recently
proposed  \cite{fastcars} which uses sequential femtosecond pulses
whereby \emph{maximal} coherence is created via adaptive
algorithms with one set of lasers and then the medium is probed by
another laser. The technique called FAST (for Femtosecond Adaptive
Spectroscopic Techniques) CARS has been applied in preliminary
experiments \cite{Beadie} and further promises to increase the
CARS Raman signal. Experimental work is underway at several
laboratories. The method offers hope for developing a way of
detecting (in real time) dangerous biological spores like anthrax.

For any spectroscopic method its relevancy and effectiveness is
eventually defined by the signal/noise ratio.  A brief review
through the literature \cite{Laserna,Eesley} shows that there have
been methods developed to overcome almost every possible
laboratory source of noise be it be unstable lasers or
non-resonant background signals, perhaps not all with the same
technique but the point is that such techniques exist.  But at the
end there is still the inherent quantum noise of the system which
is unavoidable. There are two sources for this.  The first is
truly fundamental in the sense that it represents the lower limit
of signal detection, this is the shot noise which is associated
with Poisson statistics of lasers used far above threshold. The
second arises from the spontaneous emission and the decoherence
associated with excited atoms and molecules created during all
Raman scattering processes. Often this second source of noise,
that we will  refer to as medium noise, is much larger than the
shot noise and therefore is the limiting quantum noise.

Thus our goal in this paper is to undertake a comprehensive study
using general Langevin methods to analyze the quantum noise in
CARS and FAST CARS.  Our approach and considerations have the
advantage of being general and may easily be used to study other
coherent Raman processes. We study several experimental scenarios
allowing for interactions within or without a cavity and for
pulsed or continuous wave input fields. We find a remarkable
similarity in behavior of all these cases suggesting that our
results probably have a broader validity beyond the configurations
we consider. In particular, we find that the medium (solvent)
noise is not important for FAST CARS whose efficiency is limited
by the intrinsic quantum character of the photon.

In sections II and III we define the problem in terms third order
non-linear interactions and  Langevin equations, and in section IV
we describe our model in detail, laying out the equations and
assumptions that we use. In Sec. V we consider the experiments
using sequential femtosecond pulses and in Sec. VI we discuss
experiments involving concurrent continuous wave input fields.
Then in Sec. VII we present a detailed discussion of our results
and their implications, and we provide numerical estimates based
on realistic experimental parameters.

\begin{figure}
\includegraphics*[width=\columnwidth,angle=0]{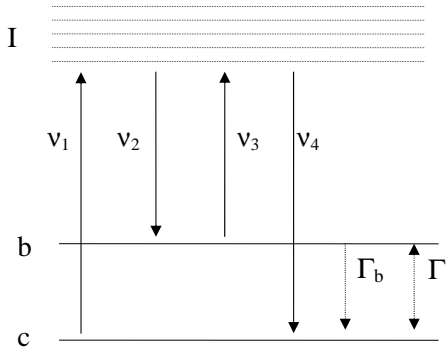}\vspace{-6cm}
\caption{CARS scheme with three input fields of frequency $\nu_1$,
$\nu_2$ and $\nu_3$ and a generated signal field
$\nu_4=\nu_1+\nu_3-\nu_2$. The radiative damping rate from level
$b$ to level $c$ is $\Gamma_b$ and the dephasing rate of the
coherence between levels $b$ and $c$ is $\Gamma$.} \label{Fig1}
\end{figure}

\section{Nonlinear Interaction}

We will consider stimulated Raman scattering involving three input
fields ${\hat a}_1,{\hat a}_2,{\hat a}_3$ and one generated field
${\hat a}_4$ with corresponding frequencies $\nu_{i=1,2,3,4}$. The
interaction Hamiltonian is written in terms of the two
Raman-coupled atomic or molecular levels $|b\rangle$ and
$|c\rangle$, of energies $\hbar\omega_b$ and $\hbar\omega_c$, and
lowering operator $\hat\sigma=|c\rangle\langle b|$
\bn
\hat{V}=\hbar\left(R_{12}\hat{a}_1\hat{a}_{2}^{\dagger}+R_{43}\hat{a}_3^{\dagger}
\hat{a}_{4}\right)\hat\sigma^{\dagger}e^{it \Delta }+
H.a.\label{Hamiltonian}\en
\emph{Here and elsewhere the time arguments will be suppressed
where obvious in order to reduce clutter}. The exponential factor
signifies that all operators are in the interaction picture, its
argument $\Delta=\omega_{bc}+\nu_{21}=\omega_{bc}+\nu_{34}$ with
$\omega_{bc}=\omega_{b}-\omega_{c}$. These particular combination
of atomic and field frequencies in $\Delta$ implies that the
Hamiltonian contains only  \emph{Raman resonant} terms and
excludes all non-resonant terms. For the case of FAST CARS where
$\nu_1$ and $\nu_2$ pulses interact with the atoms before $\nu_3$,
the non-resonant terms are simply not present. However when all
the fields are present simultaneously the permutations of the
fields lead to $24$ terms in the perturbation expansion which all
contribute at the same order \cite{Laserna}; in that case we have
in effect neglected the $20$ remaining terms which have
non-resonant combinations of the fields and the atomic levels.  If
we are close to resonance  this is certainly justified, in fact we
will assume perfect two photon Raman resonance in this paper.

The Raman coupling constants are defined by
$R_{12}=R_{12}^{'}{\cal E}_{1}{\cal E}_{2}^{*}$
\bn\label{Raman-coup} R_{12}^{'}=\frac{-1}{\hbar^2}\sum_{j}\left[
\frac{(\vec {\mu}_{fj}\cdot\vec{\varepsilon}_1)(\vec{\mu}_{j
i}\cdot\vec{\varepsilon}_2)}{(\omega_{j c}+\nu_{2})}
+ \frac{({\vec \mu}_{fj}\cdot\vec{\varepsilon}_2)(\vec{\mu}_{j
i}\cdot\vec{\varepsilon}_1)}{(\omega_{j c}-\nu_{1})}\right]\en
and likewise for $R_{43}$.  The $\vec{\varepsilon}$ are unit
polarization of the fields, $\vec {\mu}$ the dipole moments, $j$
labels the intermediate states and  ${\cal
E}_{l}=\sqrt{(\hbar\nu_{l})/(2\epsilon_0\epsilon_pV)}$ are the
field quantization factors over volume $V$ in a medium of
dielectric constant $\epsilon_p$ taken to be about the same for
all laser fields. The decoherence rate between levels $b$ and $c$
is denoted by $\Gamma$ and the depopulation rate from level $b$ to
level $c$ by $\Gamma_b$; damping to other levels is assumed
negligible.

It will help to establish a relation with macroscopic quantities.
The classical field amplitudes are $E_{i}={\cal E}_{i}\alpha_i$
for coherent state expectations $\alpha_{i}=\langle
\hat{a}_{i}\rangle$. The third-order susceptibility is defined to
be
\bn\label{chi3}\chi^{(3)}(-\nu_4,\nu_1,-\nu_2,\nu_3)&=&-\hbar\frac{R_{43}^{'*}R_{12}^{'}}
{(\Delta-i\Gamma)}\frac{N}{V},\en
with expected population difference between the two levels
$N=N_b-N_c$. The Maxwell equation for the slowly varying amplitude
of the generated field is
\bn\label{classicaleqn} \left(\frac{\partial}{\partial
t}+v\frac{\partial}{\partial
z}\right)E_{4}=\frac{i\nu_4}{2\epsilon_0\epsilon_p}\chi^{(3)}
E_{1}E_{2}^*E_{3},\en with the velocity in the medium
$v=c/\sqrt{\epsilon_p}$.

\section{Langevin Equations of motion}

The three input fields $\nu_1,\nu_2$ and $\nu_3$ are typically
strong laser fields far above threshold, we will treat them as
\emph{classical} fields and replace
$\hat{a}_{1(2,3)}\rightarrow\alpha_{1(2,3)}$. Considerable
simplification is achieved by defining
\bn\label{defA}
\hat{A}(t)&=&\frac{R_{12}\alpha_1\alpha_{2}^{*}+R_{43}\alpha_3^{*}
\hat{a}_{4}(t)}{g}\n\\
\xi &=&\frac{R_{12}\alpha_1\alpha_2^*}{g},\h{2cm}
g=|R_{43}\alpha_3^*|.\en
The normalization $g$ has been chosen such that $\hat{A}(t)$
satisfies the boson commutation rules. Thus we can treat
$\hat{A}(t)$ like a photon field operator. The parameters $g$ and
$\xi$ being dependent only on the input fields will treated as
\emph{constants} in our calculations.
The Heisenberg-Langevin equations of motion for the atomic
operators are obtained from the interaction Hamiltionian:
\bn\label{microeqns}
\dot{\hat{\sigma}}&=&-(\Gamma+i\Delta)\hat{\sigma}+ig[\hat{\sigma}_{b}
-\hat{\sigma}_{c}]\hat{A}+\hat{F}_{\sigma}\n\\
\dot{\hat{\sigma}}_{b}&=&
-\Gamma_{b}\hat{\sigma}_{b}+ig\left[\hat{A}^{\dagger}\hat{\sigma}-
\hat{\sigma}^\dagger \hat{A}\right]+\hat{F}_{b}\n\\
\dot{\hat{\sigma}}_{c}&=&\Gamma_{b}\hat{\sigma}_{b}-ig
\left[\hat{A}^{\dagger}\hat{\sigma}-\hat{\sigma}^\dagger
\hat{A}\right]+\hat{F}_{c}.\en
The coherence operator here differs from that in
Eq.~(\ref{Hamiltonian}) by a phase factor $\hat{\sigma}(t)
e^{-i\Delta t}\rightarrow\hat{\sigma}(t)$,  and
$\hat{\sigma}_{b(c)}$ are the population operators for the atomic
levels. The $\hat F$'s are the Langevin noise operators. They have
vanishing first moments $\langle \hat{F}(t)\rangle=0$, since the
entropy of the system cannot be lowered by noise \cite{Reichl}.
The second moments are taken to be \emph{delta}-correlated in time
corresponding to Markovian white noise
\bn\label{deltacorr} \langle \hat F_i(t)\hat
F_j(t')\rangle=2D_{ij}\ \delta(t-t').\en
The $D_{ij}$ are the diffusion coefficients in analogy with
classical Langevin equations. The diffusion coefficients
associated with Eqs.~(\ref{microeqns}) are calculated using the
generalized fluctuation-dissipation theorem \cite{Louisell} in
Appendix A.

The Langevin equation for the signal field operator is
\bn\label{eqa4} \dot{\hat{a}}_{4}&=&-\gamma
\hat{a}_{4}-iR_{43}^*\alpha_3\hat{\sigma}+\hat{F}_{a_4}.\en
Here we included an heuristic field damping rate $\gamma$ to allow
for atom-field interaction inside a cavity. From this equation, we
can construct the equation
\bn\label{eqA0}  \dot{\hat{A}}=-\gamma (\hat{A}-\xi)
-ig\hat{\sigma}.\en
The effects of a thermal heat bath that accounts for a
non-vanishing $\hat{F}_{a_4}$ are neglected by assuming low
temperature. In this paper we will also assume \emph{two photon
Raman resonance} thereby set $\Delta=0$.

We next define macroscopic variables by summing over the operators
for all the particles (atoms/molecules) in the medium
\bn\label{macrodef} \hat{M}=-i\sum\hat{\sigma}\h{1cm}
\hat{N}_{b(c)}=\sum\hat{\sigma}_{b(c)}\en
and the population sum and difference operators
$\hat{N}_T=\hat{N}_b+\hat{N}_c$ and $\hat{N}=\hat{N}_b-\hat{N}_c$.
The equations of motion for the collective atomic variables and
the field are then
\bn \dot{\hat{M}}&=&-\Gamma\hat{M}+g\hat{N}A+\hat{F}_{M}\n\\
\dot{\hat{N}}&=& -\Gamma_{b}(\hat{N}+\hat{N}_T)-2g[\hat{A}^{*}\hat{M}+\hat{M}^{*}\hat{A}]+{\hat{F}}_{N}\n\\
\dot{\hat{N}}_T&=&0\n\\
\dot{\hat{A}}&=&-\gamma (\hat{A}-\xi)+g\hat{M} , \en
with the total number of active particles
$\langle\hat{N}_{T}\rangle$ being conserved. The associated
diffusion coefficients are shown in Appendix~\ref{app-diff}.

Solution of these equations is facilitated by transforming them
into equivalent c-number equations \cite{Kolobov,Fleischhauer,SZ}.
This is achieved by putting all the operators in normal order
chosen to be
$\hat{A}^{\dagger},\hat{M}^\dagger,\hat{N},\hat{M},\hat{A}$ which
establishes an unique correspondence between the quantum and
classical equations. Since the equations are already in normal
order the operators are simply replaced with their classical
counterparts ${\cal A}^{*},{\cal M}^*,{\cal N},{\cal M},{\cal A}$.

 The c-number noise functions satisfy $\langle {\cal
F}(t)\rangle=0$ and $\langle {\cal F}(t){\cal F}(t')\rangle=2{\cal
D}\delta(t-t')$ just like their operator counterparts.
The c-number diffusion coefficients will however acquire
additional terms arising from normal ordering;  all the
non-vanishing coefficients  are listed in Appendix~\ref{app-diff}.
The corresponding properties of the noise in the frequency domain
are summarized in Appendix~\ref{app-spectrum}. It is worth
mentioning here that the expectations in the c-number
representation are now in the Glauber-Sudarshan P-representation
of a thermal distribution which happens to be a Gaussian
distribution of zero mean. This means that the Gaussian moment
theorem applies and the first and second moments suffice to
determine the distribution completely.

\section{Equations and assumptions of the
model}\label{sec-eqn-assum}

We consider small fluctuations about the mean values for both
atomic and field variables ${\cal M}(t)={\cal M}_0(t)+\delta{\cal
M}(t)$, ${\cal N}(t)={\cal N}_0(t)+\delta{\cal N}(t)$ and ${\cal
A}(t)={\cal A}_0(t)+\delta{\cal A}(t)$. It is convenient to define
two real variables ${\cal T}(t)={\cal M}^*(t){\cal A}_0(t)+{\cal
M}(t){\cal A}_0^*(t)$ and ${\cal S}(t)=|{\cal A}(t)|^2$, with
linearized fluctuations
\bn
\delta {\cal S}(t)&=&{\cal A}_{0}(t)\delta A^*(t)+{\cal A}_{0}^*(t)\delta A(t)\n\\
\delta {\cal T}(t)&=&{\cal A}_{0}(t)\delta M^*(t)+{\cal
A}_{0}^*(t)\delta M(t)\n\\
{\cal F_T}(t)&=&{\cal A}_{0}(t){\cal F_M}^*(t)+{\cal
A}_{0}^*(t){\cal F}_{{\cal M}}(t).\en
The relation to the properties of the generated field $\alpha_4$
that we are finally interested in are as follows: The strength of
the signal field, given by the number of generated photons, and
the atomic noise in the photon number, quantified by its
\emph{variance}, are determined by
\bn n_4(t)&=&\langle\alpha_4^*(t)\alpha(t)\rangle=|{\cal
A}_0(t)-\xi|^2\n\\
\delta n_4^2(t)&=&\frac{n_4(t)}{|{\cal A}_0(t)|^2}\langle\delta
{\cal S}^2(t)\rangle.\en
To reduce clutter we will often leave out the expectation brackets
being obvious from the context. The variance is obviously
linearized and neglects relatively small terms quadratic in the
correlations. We will primarily consider regimes where the
generated field $\alpha_4$ is much weaker than the input fields
(as we justify numerically in Sec.~\ref{Sec:signal}) and therefore
we will replace $|{\cal A}_0|^2\simeq |\xi|^2$ where appropriate
and consistent.

\subsection{adiabatic elimination of the averages of atomic variables}

Since the fluctuations have vanishing average, we can write
separate equations for the mean and the linear fluctuations. The
equations for the averages can be had by simply leaving out the
Langevin forces
\bn\label{cnumeqn} \partial_
t{\cal M}_0&=&-\Gamma{\cal M}_0+g{\cal N}_0{\cal A}_0\n\\
\partial_ t{\cal N}_0&=& -\Gamma_{b}({\cal N}_0+N_T)-2g[{\cal
A}_0^{*}{\cal M}_0
+{\cal M}_0^{*}{\cal A}_0]\n\\
\partial_ t{\cal A}_0&=&-\gamma({\cal
A}_0-\xi)+g{\cal M}_0. \en
Atomic/molecular transitions typically occur on a faster time
scale than the variations in a radiation field, which means that
the medium will adiabatically follow the field for weak coupling.
Even when using fast pulses, the coherence generating pulses can
be taken to be of sufficiently long duration for this to be true.
Therefore taking averages over intermediate time scales, the
time-derivatives in the equations for the averages of the medium
variables ${\cal M}_0$ and ${\cal N}_0$ may be neglected in what
is called adiabatic elimination \cite{Louisell} and we obtain
algebraic relations for the coarse-grained averages:
\bn\label{means} {\cal N}_0=-\frac{N_T}{1+B|\xi|^2},\ \ {\cal
M}_{0}= -\frac{C}{g}\frac{{\cal A}_0}{1+B|\xi|^2}\en
where we set $|{\cal A}_0|^2\simeq |\xi|^2$ as discussed earlier.
We define some parameters here that we will use often and which
will serve to simplify our expressions considerably:
\bn
B=\frac{4g^2}{\Gamma_b\Gamma},\h{1cm}C=\frac{N_Tg^2}{\Gamma}.\en
The parameter $B$ is dimensionless and $C$ carries the dimension
of inverse time.  As we corroborate numerically later $B|\xi|^2\ll
1$, so the upper level is never strongly populated, ${\cal
N}_0\simeq -N_T$ and we can simplify by setting $1+B|\xi|^2\simeq
1$ in the rest of the paper. Substituting the atomic mean values
we get an uncoupled equation for the average of the field variable
\bn \partial_ t{\cal A}_0=-\gamma({\cal A}_0-\xi)-C{\cal A}_0
\label{mean-eqn}.\en

\subsection{fluctuations}

In describing the fluctuations, we cannot make arguments for
adiabatic elimination of the medium variables as we did for the
averages; this is due to the presence of the rapidly varying noise
functions.  An attempt to make such an approximation can lead to
unphysical divergences in the correlations; we will discuss this
issue in some detail in Appendix \ref{app-adel}. Therefore we
solve coupled equations for the fluctuations for \emph{both} the
field and the medium.  We write the relevant equations in terms of
the variables ${\cal T}$ and ${\cal S}$
\bn \partial_ t\delta {\cal T}&=& -\Gamma \delta{\cal
T}+2g\delta{\cal N}|{\cal A}_0|^2+g{\cal N}_0\delta{\cal S}
+{\cal F_T}\n\\
\partial_ t\delta{\cal N}&=& -\Gamma_b\delta{\cal
N}+2C \delta {\cal S}  -2g\delta {\cal T}
+{\cal F}_{\cal N}\n\\
\partial_ t\delta{\cal S}&=&-\gamma\delta{\cal S}
+g\delta{\cal T}. \label{fluc-eqns}\en
In writing these equations, we have treated ${\cal A}_0$ as a
constant.  We will see that in all the cases we consider in this
paper this assumption is valid, because either ${\cal A}_0$ will
have an unvarying steady state value or it will satisfy ${\cal
A}_0\simeq \xi$, even as a function of position when propagation
in free space is inovlved (see next subsection).  It is clear from
the equations above that the fluctuations can be fully expressed
in terms of the correlations of the variables ${\cal T}$ and $\cal
N$. Using the results in Appendix~\ref{app-diff}, we find that the
two variables are uncorrelated with each other but the strengths
of their autocorrelations are specified by the diffusion
coefficients
\bn\label{diff-coeff} 2{\cal D}_{{\cal T}{\cal
T}}&=&2(\Gamma-\Gamma_b)N_TB|\xi|^4\n\\
2{\cal D}_{{\cal N}{\cal N}}&=&4\Gamma_bN_T B|\xi|^2\en
where we used the expressions for the averages in
Eq.~(\ref{means}). In writing these expressions we used the weak
field approximation ${\cal A}_0\simeq\xi$ and the assumption of
weak upper level occupation $B|\xi|^2\ll 1$, and therefore in
effect we can treat these coefficients as constants.

\subsection{Free space description}

The equations above are appropriate for describing experiments
conducted in a cavity of damping rate $\gamma$. However if we wish
to describe experiments in free space we have to allow for field
propagation through the medium and effectively replace
$\gamma\rightarrow v\partial_z$ . An approach discussed by
Drummond and Carter \cite{Drummond,Fleischhauer} outlined in
Appendix \ref{app-prop}, leads to the appropriate space-time
equation for the field variable
\bn (\partial_ t+v\partial_z){\cal A}(z,t)=g{\cal M}(z,t). \en
This equation follows from Eq.~(\ref{spacetime}) derived in the
appendix, on assuming that all the input fields propagate
collinearly in the z-direction and their amplitudes
$\alpha_{1(2,3)}$ vary little over the interaction length $L$,
i.e. the length of the region over which the fields interact with
the atoms, and therefore may be treated as constants.

Thus for experiments in free space, equations for the field
variable in (\ref{mean-eqn}) and (\ref{fluc-eqns})  are modified
to
\bn (\partial_ t+v\partial_z){\cal A}_0(z,t)&=&-C {\cal A}_0(z,t)\n\\
(\partial_ t+v\partial_z)\delta{\cal S}(z,t)&=& g\delta{\cal
T}(z,t). \label{free-space}\en
The equations for the atomic variables are formally unchanged, but
they now carry both position and time argument $(z,t)$; their
spatial dependence arises through their interdependence on the
field variable.

\section{Pulsed Inputs}

We first consider the case where the input laser fields are
extremely short and fast pulses.  In particular we take the pulses
to arrive in sequence as is the case in FAST CARS \cite{fastcars};
the fields $\alpha_1$ and $\alpha_2$ are allowed to interact first
with the atoms for a duration $\Delta t_c$ creating the coherence,
then after a delay $\Delta t_d$  the field $\alpha_3$ occurs for
time $\Delta t_3$. This means that the atomic variables can now
only depend on the first two fields, so we replace ${\cal
A}_0\rightarrow \xi$ in Eqs.~(\ref{mean-eqn}) and
(\ref{diff-coeff}) and also set  $\delta{\cal S}=0$ in the
equations for the fluctuations of atomic variables:
\bn \partial_ t\delta {\cal T}&=& -\Gamma \delta{\cal
T}+2g\delta{\cal N}|\xi|^2
+{\cal F_T}\n\\
\partial_ t\delta{\cal N}&=& -\Gamma_b\delta{\cal
N}  -2g\delta {\cal T} +{\cal F}_{\cal N}.\label{fluc-eqns-p}\en

The main premise for this scenario is that when a rapid sequence
of fast femtosecond pulses are used, the field $\alpha_3$ scatters
off the generated coherence before it has time to decay
significantly. Thus for time intervals such that $\Delta
t_d+\Delta t_3\ll\Gamma^{-1},\Gamma_b^{-1}$ the generated
coherence may be taken to be constant.  For longer durations
however we have to allow for decay of the coherence ${\cal M}_0$
in Eq.~(\ref{means}).

During the signal-generating cycle when  $\alpha_3$ is present the
mean coherence is described by
\bn \partial_ t{\cal M}_0&=& -\Gamma{\cal M}_0 -g({\cal
A}_0-\xi){\cal N}_0.\en
After the coherence generating pulses are turned off, the value of
generated coherence from Eq.~(\ref{means}) ${\cal M}_0(0)= g\xi
{\cal N}_0/\Gamma$ can be taken as the initial value of the
coherence in the above equation (if $\Delta t_d\simeq 0$)  and
since initially there is no signal field ${\cal A}_0(0)=\xi$. It
is then easy to see that the ratio of the magnitude of the first
term to that of the second term is $\sim |\xi|/|{\cal A}_0-\xi|$
which is large for weak generated field and short duration pulses.
Therefore we can neglect the second term and take the driving
coherence to simply decay exponentially from the moment the
generating fields are turned off ${\cal M}_0(t)={\cal
M}_0(0)e^{-\Gamma t}$.

The equation for the average field then becomes
\bn \partial_ t{\cal A}_0=-\gamma({\cal A}_0-\xi)-C \xi e^{-\Gamma
t}\label{mean-eqn-1}.\en
The fluctuations will likewise decay since the diffusion
coefficients depend on the averages ${\cal N}_0$ and ${\cal M}_0$.
The diffusion coefficients in Eq.~(\ref{diff-coeff}) will then
have an exponentially decaying time dependence, ${\cal D_{NN}}\sim
e^{-\Gamma_b t}$ and  ${\cal D_{TT}}\sim e^{-\Gamma t}$. We note
however that there is no constraint on the duration $\Delta t_c$
of the coherence generating pulses, so they can be taken to be
sufficiently long to create a steady state value of the coherence
before they are turned off, thereby justifying adiabatic
elimination.

\subsection{pules: interaction in a cavity}\label{Sec:intcav}

The mean value of the field variable after the third pulse ${\cal
A}_0(\Delta t_3)$ is easily obtained by integrating
Eq.~(\ref{mean-eqn-1})
\bn {\cal A}_0(\Delta t_3)-\xi= -C \xi\times \frac{e^{-\Gamma
\Delta t_3}-e^{-\gamma \Delta t_3}}{\gamma-\Gamma}. \en
The phase of ${\cal A}_0$ is given by that of $\xi$, since ${\cal
A}_0/\xi$ is real.  When the duration of the third pulse is short
$\Delta t_3\ll \Gamma^{-1}$,
\bn {\cal A}_0(\Delta t_3)-\xi\simeq -C \xi\Delta t_3.\en
The equal time fluctuations of the field are given by
\bn  \delta{\cal S}^2(\Delta t_3)=g^2\int_0^{\Delta
t_3}\h{-5mm}dt\int_0^{\Delta t_3}\h{-5mm}dt'e^{-\gamma (2\Delta
t_3-t-t')}\langle\delta{\cal T}(t)\delta{\cal
T}(t')\rangle\label{fluct-pc}\en
In order to evaluate this we first take a Fourier transform of the
Eqs.~(\ref{fluc-eqns-p}), the relevant definitions and properties
of the fluctuations and noise in the frequency domain are
described in Appendix~\ref{app-spectrum}:
\bn (\Gamma-i\omega)\delta{\cal T}(\omega)&=& 2g|\xi|^2\delta{\cal
N}(\omega)+{\cal F_T}(\omega)
\n\\
(\Gamma_b-i\omega)\delta{\cal N}(\omega)&=&  -2g\delta {\cal
T}(\omega)+{\cal F}_{\cal N}(\omega)\label{fluct-eqns-p-freq}\en
which we then solve to get
\bn \delta{\cal T}(\omega) \simeq
\frac{\left[2g|\xi|^2{\cal F}_{{\cal N}}+(\Gamma_b-i\omega){\cal
F_T}\right]}{(\Gamma-i\omega)(\Gamma_b-i\omega)}. \label{eqnT}\en
Assuming weak excitation we set
$|1+\frac{4g^2\xi^2}{[\Gamma-i\omega] [\Gamma_b-i\omega]}|^2\simeq
1 $.  Using this expression for $\delta {\cal T}(\omega)$ we can
evaluate the integrals in Eq.~(\ref{fluct-pc}). The details of the
calculations are shown in Appendix~\ref{app-nonad}.  The general
result result derived there is not particularly illuminating,
instead here we consider the relevant limiting cases. In the short
pulse limit $\Delta t_3\ll \Gamma^{-1}$, we found that a Taylor
expansion leads to an exact cancellation of the terms linear in
$\Delta t_3$ and we get a quadratic dependence on the interaction
time
\bn \delta{\cal S}^2(\Delta t_3) \simeq \frac{g^2\Delta
t_3^2}{2}\left[ \frac{B|\xi|^42{\cal D_{NN}}}{(\Gamma+\Gamma_b)}
+\frac{2{\cal D_{TT}}}{\Gamma} \right].\label{pulse-short}\en
A conspicuous feature of this limit is that $\gamma$ is absent in
the expressions for both the mean and the fluctuation. Thus for
pulses of duration shorter than the cavity damping time, the
presence of the cavity has no effect on the signal or its
associated noise due to the medium. This short pulse limit implies
the hierarchy of time scales
$\gamma^{-1}\gg\Gamma^{-1},\Gamma_b^{-1}\gg\Delta t_3$.

In the opposite limit of long pulse duration $\Delta t_3\gg
\gamma^{-1}\gg\Gamma^{-1},\Gamma_b^{-1}$ the fluctuation is
approximately
\bn \delta{\cal S}^2(\Delta t_3)\simeq
 \frac{g^2\Delta
t_3^2e^{-\Gamma \Delta t_3}}{2}\left[\frac{2 {\cal
D_{NN}}}{\Gamma\Gamma_b\gamma}B|\xi|^4
 +\frac{2{\cal
D_{TT}}}{\Gamma^2\gamma}\right].\label{pulse-long}\en
Here we set the depopulation rate and decoherence rates to be
equal $\Gamma\simeq \Gamma_b$, to mask unnecessary details and
highlight the main feature, which is that both the mean and the
fluctuations essentially vanish towards the end of a long third
pulse as the driving coherence disappears. The conclusion is that,
regardless of the duration of the third pulse, significant signal
field is generated only during times satisfying
$<\Gamma^{-1},\Gamma_b^{-1}$.

\subsection{pulses: interaction in free space}

The last subsection showed that increasing the duration of the
interaction between the third pulse and the atoms will make a
difference only upto a point, since we are limited by the decay
time of the coherence. In free space, this means that we will not
get a stronger sustained signal by simply increasing the
interaction length L. Therefore we will confine ourselves to
sample sizes $L<v\Gamma^{-1},v\Gamma_b^{-1}$, and we will use
Eq.~(\ref{free-space}) where, taking the coherence to remain
unchanged during the time of interaction, we set ${\cal
A}_0(z,t)\rightarrow \xi$.

A time-frequency Fourier transform and subsequent integration over
the interaction length gives
\bn (-i\omega+v\partial_z){\cal
A}_0(z,\omega)&=&-C \xi\delta(\omega)\\
\Rightarrow {\cal A}_0(L,\omega)&=&e^{i\omega L/v}\xi\left[1-C
\frac{L}{v}\right]\n\en
and an inverse transform yields the delta-function $\delta(t-L/v)$
that simply tells us that time is a redundant parameter and the
field can be specified by its position alone
\bn {\cal A}_0(L)&=&\xi-C \xi\frac{L}{v}.\en

Likewise for the variance we consider the equations for the
fluctuations in the frequency domain.  Since we set  ${\cal
A}_0(z,t)\simeq \xi$, the equations are given simply by
(\ref{fluct-eqns-p-freq}).  Substituting the expression
$\delta{\cal T}(\omega)$ from Eq.~(\ref{eqnT}) into the equation
for the field fluctuation
\bn-i\omega+v\partial_z\delta{\cal S}(z,\omega)= g\delta{\cal
T}(z,\omega),\en
and integrating over the interaction length $L$ we get the
spectral density of the noise
\bn\delta S^2(L,\omega)= \frac{1}{2\pi}\frac{L^2}{v^2}
\frac{\left[4g^4|\xi|^4 2{\cal D_{NN}}+g^2
 (\Gamma_b^2+\omega^2)2{\cal
D_{TT}}\right]}{(\Gamma^2+\omega^2)(\Gamma_b^2+\omega^2)}.\n\en
The autocorrelation at equal times obtained by doing a partial
fraction decomposition and an inverse Fourier transform defines
the variance
\bn \delta S(L)^2 =\frac{g^2L^2}{2v^2}\left[ \frac{B|\xi|^4 2{\cal
D_{NN}}}{(\Gamma+\Gamma_b)} +\frac{2{\cal D_{TT}}}{\Gamma}
\right].\en
We see that both the mean and the variance are identical to what
we found in the short pulse limit when the interaction took place
inside a cavity, bearing in mind that here $L/v$ defines the time
of interaction. This reaffirms the conclusion that for short
pulses there is no real advantage in using a cavity.

\section{Continuous Wave (cw) Inputs}\label{Sec:freespace}

We now consider the cases where all three input fields occur
continuously and simultaneously.  In this case steady state values
for the medium variables may be considered, and their decay does
not put limits on the interaction time as it did previously when
using sequential pulses. But the calculation of  fluctuations is
complicated by the fact that the equations for the medium
variables and the field variable do \emph{not} decouple as they
did for sequential pulses.

\subsection{cw: interaction in a cavity}

The average is found by integrating Eq.~(\ref{mean-eqn}), exactly
in the form it is written, with the initial condition ${\cal
A}_0(0)=\xi$,
\bn {\cal A}_0(t)-\xi= -C \xi\times\frac{1-e^{-(\gamma+C )
t}}{(\gamma+C )}.\en
Unlike the pulsed case there are no obvious time constraints,
except those arising from possible damage to the sample by
prolonged exposure to the fields, so we can take the long time
limit $t\rightarrow\infty$ and we get a steady state signal
\bn {\cal A}_0-\xi= -C \xi\times\frac{1}{(\gamma+C )}.\en
We note that if $\gamma\ll C $ the generated field essentially
vanishes, which suggests that we need to have $\gamma\geq C $. In
that case we can further use the assumption of weak signal field
to conclude that
\bn \frac{{\cal A}_0-\xi}{\xi}\simeq \frac{C }{\gamma}\ll 1
\h{5mm}\Rightarrow {\cal A}_0-\xi\simeq -C
\xi\frac{1}{\gamma}.\label{C-approx}\en

The fluctuations are best determined by writing the three coupled
equations in the frequency domain in matrix form:
\bn \left[\begin{array}{ccc}
(\Gamma-i\omega)&- 2g|{\cal A}_0|^2 & -g{\cal N}_0\\
2g &(\Gamma_b-i\omega)& -2C \\
-g & 0 & (\gamma-i\omega)\end{array}\right]
\left[\begin{array}{c}\delta{\cal T}\\ \delta{\cal N}\\
\delta{\cal S}\end{array}\right]
=\left[\begin{array}{c}{\cal F}_{{\cal T}}\\ {\cal F}_{{\cal N}}\\
0\end{array}\right].\n\en
Inverting the coefficient matrix ${\bf M_F}$ gives the solution
\bn \delta{\cal S}(\omega)&=&\frac{\left[(\Gamma_b-i\omega)g{\cal
F_T}(\omega)+
 2g^2|{\cal A}_0|^2{\cal F_N}(\omega)\right]}{\det{\bf M_F}} \\
 \det{\bf M_F}
&\simeq&(\Gamma-i\omega)(\Gamma_b-i\omega)(\gamma-i\omega).\n\en
In computing the determinant in the denominator we introduced some
simplifications using the arguments leading upto
Eq.~(\ref{C-approx}) and we neglected the frequency dependence in
the term proportional to $B|{\cal A}_0|^2$ [refer to the comment
following Eq.~(\ref{eqnT})] . Squaring this gives the power
spectrum (Appendix~\ref{app-spectrum}) and then a Fourier
transform and the adiabatic assumption $\gamma\ll \Gamma,\Gamma_b$
yields the steady state fluctuations
\bn \delta{\cal S}^2(t)\simeq\frac{g^2}{2\gamma}\left[\frac{2
{\cal D_{NN}}}{\Gamma\Gamma_b}B|\xi|^4
 +\frac{2{\cal D_{TT}}}{\Gamma^2}\right].\label{cw-long}\en
While the adiabatic assumption was not necessary it gave a simpler
expression, the more general expression is shown in
Appendix~\ref{app-nonad}. What sets this case apart from the rest
is that both the signal field and the variance in the steady state
depend on the cavity lifetime, $\gamma^{-1}$ and that dependence
is linear.

\subsection{cw: interaction in free space}

In order to find the signal strength, we do a Fourier transform of
Eq~(\ref{free-space})
\bn (-i\omega+v\partial_z){\cal A}_0(z,\omega)&=&-C {\cal
A}_0(z,\omega)\en
and integrate it over the interaction length $L$ to get
\bn {\cal A}_0(L,\omega)&=&e^{i\omega L/v}{\cal
A}_0(0,\omega)e^{-C L/v}.\en
As in the analogous case in a cavity, the time parameter is seen
to be redundant after an inverse  Fourier transform and we get
simply
\bn {\cal A}_0(L)&=&\xi e^{-C L/v}.\label{cwfs-mean}\en

In calculating the fluctuations we take the Fourier transform of
the equations for the atomic variables as they appear in
Eqs.~(\ref{fluc-eqns}) and the field fluctuation as it appears in
(\ref{free-space}), keeping in mind that all the fluctuation and
noise elements will thereafter carry an argument of $(z,\omega)$.
The three coupled equations yield an equation for the field
fluctuation
\bn\label{cont-fs} [v\partial_z-i\omega-{\cal
G}(\omega)]\delta{\cal S}\simeq
\frac{g\left[2g|\xi|^2{\cal F}_{{\cal N}}+(\Gamma_b-i\omega){\cal
F_T}\right]}{(\Gamma-i\omega)(\Gamma_b-i\omega)} \n\\
{\rm with}\ \ {\cal G}(\omega)\simeq \frac{g\left[4gC
|\xi|^2+(\Gamma_b-i\omega)g{\cal N}_0\right]
}{(\Gamma-i\omega)(\Gamma_b-i\omega)}.\h{5mm} \en
We integrate this with respect to the spatial coordinate $z$, and
thereby we get the spectral density
\bn \delta S^2(L,\omega)&=&\frac{1}{2\pi}\frac{L}{v}\times
\frac{1-e^{2Re[{\cal G}(\omega)]\frac{L}{v}}}{-2Re[{\cal G}(\omega)]}\n\\
& &\times\frac{4g^4|\xi|^4 2{\cal D_{NN}}+g^2
 (\Gamma_b^2+\omega^2)2{\cal
D_{TT}}}{(\Gamma^2+\omega^2)(\Gamma_b^2+\omega^2)}.\n\en
We use $B|\xi|^2\ll 1$ to simplify
\bn 2Re[{\cal G}(\omega)] &\simeq& -\frac{2C
\Gamma^2}{(\Gamma^2+\omega^2)}. \en
We note this quantity achieves the largest magnitude when
$\omega=0$.  In the case of short interaction length $C L/v\ll 1$
both the signal and the variance resemble those we got in the case
of pulsed inputs in free space
\bn {\cal A}_0(L)&\simeq& \xi-C \xi\frac{L}{v}\\
\delta S^2(L,\omega)& \simeq& \frac{1}{2\pi}\frac{L^2}{v^2}
\frac{\left[4g^4|\xi|^4 2{\cal D_{NN}}+g^2
 (\Gamma_b^2+\omega^2)2{\cal
D_{TT}}\right]}{(\Gamma^2+\omega^2)(\Gamma_b^2+\omega^2)}.\n\en

In the opposite time limit of long interaction length $C L/v\gg
1$, we find that the expression for the average value of the field
variable ${\cal A}_0$ tends to become increasingly smaller.
Equation (\ref{cwfs-mean}) tells us
\bn n_4(L)=|{\cal A}_0(L)-\xi|^2&=&|\xi(1- e^{-C L/v})|^2,\en
so that in this limit the input fields vanish and give way to the
signal field. Our model however does not allow us to obtain an
accurate expression for the variance in this limit. This is
because the equations (\ref{fluc-eqns}) were based on the
assumption that ${\cal A}_0$ does not change significantly and the
generated field is relatively weak, and this is no longer true
when $L$ becomes large.  Yet based on the fact that average value
${\cal A}_0$ decreases over the length $L$ and hence the diffusion
coefficients also decrease, we could expect that the noise
contribution from the active medium will actually be lower farther
along the interaction length.

\section{Discussion of Results, Inferences and Assumptions}

We will now discuss in some detail the physical implications of
the results we obtained in the previous sections, and also
elaborate a bit more on our assumptions and some associated
subtleties that we touched on while deriving those results. We
begin by considering realistic numerical values for our parameters
and variables and thereby provide concrete justification for the
approximations we made.

\subsection{Numerical Estimates}

For the purpose of numerical estimation we will use the specific
example of an anthrax spore for which the Raman active molecule is
dipicolinic acid (DPA) which constitutes $17\%$ of the weight, the
rest being mainly water.  The number density of DPA molecules in
an anthrax spore is $N_T\sim 4\times 10^{26}$ molecules m$^{-3}$,
and the dimensions of the spore itself is $1\times 2\times 1\ {\rm
\mu m}^3$.   We will take all optical frequencies to be in the
visible range $\nu\sim 2\pi c/\lambda\simeq 4\times 10^{15}$.

Since some of the properties of DPA are not easily available, for
those properties we will use the values for benzene, an organic
molecule that has a similar structure. Thus for instance we use
the spontaneous Raman differential cross-section for Benzene
$\frac{d\sigma}{d\Omega}=32.5\times 10^{-34} {\rm
m^2/sr/molecule}$, and we take the linewidths of Raman transitions
in Benzene to estimate $\Gamma_b\sim 10^{11}{\rm Hz}$.

We first determine the strength of the Raman coupling $R'_{ij}$,
we can do that in two ways, assuming in either case that only a
few intermediate states contribute: first using
Eq.~(\ref{Raman-coup}), and taking $\mu\sim e a_0$ with $a_0$
being the Bohr radius
\bn |R'_{ij}|\sim \frac{\mu^2}{\hbar^2\nu}\simeq 4\times
10^{-6}{\rm C}^2{\rm m}^2/{\rm J}^2{\rm s}; \en
and secondly using the Kramers-Heisenberg formula for the
differential scattering cross-section applied to spontaneous Raman
scattering
\bn |R'_{ij}|\sim\frac{4\pi
\epsilon_0c^2}{\hbar\nu^2}\sqrt{\frac{d\sigma}{d\Omega}}\simeq
0.35\times10^{-6} {\rm C}^2{\rm m}^2/{\rm J}^2{\rm s}.\en
So we will take $|R'_{ij}|\sim 10^{-6}\ {\rm C}^2{\rm m}^2/{\rm
J}^2{\rm s}.$  Next we determine the field density and photon
density from the expression for the field intensity (power/area)
\bn I=\frac{1}{2}v\epsilon_p\epsilon_{0}|{\cal
E}\alpha|^2=\frac{1}{4}v(\hbar\nu)\frac{|\alpha|^2}{V}.\label{power}\en
For strong lasers we could take the typical intensities to be
$I\sim 10^{12}$ Wm$^{-2}$, in which case we get
\bn |{\cal E}\alpha|^2\sim 5\times 10^{14}{\rm N}^2{\rm C}^{-2},
\h{3mm} \frac{n}{V}=\frac{|\alpha|^2}{V}\simeq 5\times 10^{22}{\rm
m}^{-3}.\en
Taking all the input fields to have similar intensities and
assuming natural linewidth $\Gamma_b\sim 10^{11}$ Hz and
decoherence rate $\Gamma\sim 10^{13}-10^{14}$ Hz,
\bn
B|\xi|^2=\frac{4g^2}{\Gamma\Gamma_b}\sim\frac{4|R'_{12}|^2|{\cal
E}\alpha|^2}{\Gamma\Gamma_b}\sim 10^{-8}.\en
Indeed for this choice of parameters  we are justified in taking
the weak-excitation limit $B|\xi|^2\ll 1$, and in fact this will
be valid till the field intensities increase to about $10^{16}$
Wm$^{-2}$; we note that this is the intensity level at which
cascade breakdown of air occurs at STP \cite{Kroll}, so our
weak-field assumption covers most realistic regimes.  Thus we are
quite justified in setting $1+B|\xi|^2\simeq 1$.

Finally we address the issue of the interaction times, since many
of our results assume short interaction times.  In the case of
pulsed inputs, $\Gamma^{-1}$ sets the limits on the pulse duration
to be  $\Delta t_3< 10^{-13}$ s, which is certainly within the
bounds of experimental capabilities using femtosecond pulses.  In
the case of propagation in free space, taking the dimension of an
anthrax spore $L\sim 10^{-6}$ m as the interaction length we find
$L/v\sim 10^{-15}<\Gamma^{-1}$.  In either case the constraints of
short interaction times are likely to be satisfied for physical
regimes of interest.

\subsection{Signal and noise}\label{Sec:signal}

The main observation we have from our calculations is that when
the duration of interaction between the signal-generating field
and the Raman-active medium $\Delta t=\Delta t_3,L/v$ is short
compared to $\Gamma^{-1}$, the signal and the noise due to the
medium have  essentially the same theoretical description
independent of the various experimental scenarios that we
considered:
\bn n_4&\simeq&\xi^2C ^2\Delta t^2\simeq\frac{
N_T^2|R_{43}R_{12}|^2n_1n_2n_3 \Delta t^2}{\Gamma^2}\n\\
 \delta n_4^2 &\simeq& \frac{g^2\Delta
t^2}{2}\left[ \frac{B|\xi|^42{\cal D_{NN}}}{(\Gamma+\Gamma_b)}
+\frac{2{\cal D_{TT}}}{\Gamma}
\right]\n\\
&\simeq& n_4^2\times \frac{8(\Gamma-\Gamma_b)}{N_T\Gamma_b},\en
where we have used $B|\xi|^2\ll 1$ to set ${\cal N}_0\simeq N_T$
and $C \simeq C$; we also noted that the term involving ${\cal
D_{NN}}$ is smaller by a factor of $B|\xi|^2$  than the one
involving ${\cal D_{TT}}$ and hence we only retained the latter.

The exception to the above expressions was the case of continuous
wave input in a cavity for which we got a steady state signal and
noise given by:
\bn n_4&\simeq&\frac{
N_T^2|R_{43}R_{12}|^2n_1n_2n_3 }{\Gamma^2}\frac{1}{\gamma^2}\n\\
\delta n_4^2&\simeq& n_4^2\times
\frac{8\gamma(\Gamma-\Gamma_b)}{N_T\Gamma\Gamma_b}.\en
The signal has the same form as the other cases with $\Delta
t\rightarrow \gamma^{-1}$. But the fluctuations differ by a factor
of $\gamma/\Gamma$.  Since typically $\gamma\ll\Gamma$ the noise
noise will be less in this case, and because $\gamma^{-1}$ is
greater than the short time limits in the other cases the signal
is bigger. In addition to this, the fact that there are no major
constraints on the duration of the irradiations apart from their
possible destruction of the sample, this scenario seems to be the
best one from the signal to noise perspective.  However that
conclusion has to be moderated by the fact that when all three
input fields happen simultaneously, there can be significant
contributions from the non-resonant terms in our interaction
Hamiltonian in Eq.~(\ref{Hamiltonian}), particularly if we cannot
achieve two-photon Raman resonance. This is where the sequential
pulse scheme as in FAST CARS has an advantage.

We can recast our expressions for the signal and fluctuations in
terms of macroscopic variables. First using Eq.~(\ref{chi3}) and
the definition of classical field amplitude $E_l={\cal
E}_l\alpha_l$ we write the signal in terms of the classical field
amplitudes:
\bn  |E_4|^2&=&
\left(\frac{\nu_{4}}{2\epsilon_0\epsilon_p}\right)^2[\chi^{(3)}]^2
|E_1|^2|E_2|^2|E_3|^2\times \Delta t^2\en
where $\Delta t=\Delta t_3,L/v,\gamma^{-1}$ depending on the
experimental configuration.  Written this way we see that our
expression for the signal is consistent with what we would get if
we integrated the classical equation (\ref{classicaleqn}) for the
slowly-varying amplitude with appropriate assumptions. A more
practical representation would be in terms of the input
intensities using Eq.~(\ref{power})
 \bn   I_4&=&
 \left(\frac{\nu_{4}}{c\epsilon_{0}^2}\right)^2
[\chi^{(3)}]^2I_1I_2I_3 \Delta t^2\n\\
\delta I_4^2
&=&I_{4}^2\times\frac{8(\Gamma-\Gamma_b)}{\Gamma_bN_T} \times
\left(\frac{\gamma}{\Gamma}\ {\rm for\ cw\ \&\ cavity}\right).\en

At this point we validate numerically the assumption of weak
signal relative to the input fields.  Using the parameters in the
previous subsection we find the third order susceptibility to be
$\chi^{(3)}\sim 10^{-34}$ C$^4$N$^{-3}$m$^{-2}$. Taking all the
input fields to have the same intensity $I_i\simeq I_1\sim I_2\sim
I_3$ and the ratio of the signal intensity to the input field
intensity is
\bn  \frac{I_4}{I_i}\simeq 10^{-10}I_i^2\Delta t^2.\en
For the short interaction times that we consider $\Delta t\simeq
10^{-13}$~s and input field intensities of $I_i\sim 10^{12}$
Wm$^{-2}$ this works out to be $I_4\simeq 10^{-12}{I_i}$.  This
shows that even for higher input field intensities and increased
density of active molecules, the signal field would still remain
relatively weak.

\subsection{Comparison with Shot Noise}

Starting with the expression we derived for the generated field,
we can write the operator for the generated field in terms of the
input fields:
\bn \hat{a}_4=\frac{ N_T|R_{43}R_{12}|\Delta
t}{\Gamma}\hat{a}_1\hat{a}_2^\dagger\hat{a}_3.\en
Then noting that
\bn a_4^\dagger a_4&\propto& n_1(n_2+1)n_3\n\\
 a_4^\dagger
a_4a_4^\dagger a_4&\propto& (n_1^2+n_1) (n_2^2+ 3n_2+1)
(n_3^2+n_3)\en
and on assuming similar and large photon numbers in the input
field, $n_1\sim n_2\sim n_3=n_i\gg 1$ we find that the variance
corresponding to the shot noise is:
\bn\delta n_{(shot)}^2\simeq \left(\frac{ N_T|R_{43}R_{12}|\Delta
t}{\Gamma}\right)^4\times 3\times n^5\simeq 3\frac{n_4^2}{n_i}\en

Therefore the ratio of the medium noise and the shot noise is:
\bn \frac{\delta n^2}{\delta n_{(shot)}^2}\simeq
\frac{8(\Gamma-\Gamma_b)}{3\Gamma_b}\times \frac{n_i/V}{N_T/V}.\en
Using the numerical values that we considered earlier we find
\bn \frac{\delta n^2}{\delta n_{(shot)}^2}\simeq
10^{-2}-10^{-1}\en
which shows that the noise due to the medium is less than the shot
noise. The shot noise increases with increasing input field
intensities while the medium noise increases with increasing
number of particles.  But in the ratio it is interesting that the
behavior is exactly the opposite, the weight of the shot noise
decreases with increasing intensity of the input fields while the
weight of the medium noise decreases with increasing density of
the medium. Therefore when the field intensity becomes very strong
for a given medium density, shot noise could become lesser, but as
we noted earlier the intensities cannot be much stronger than what
we already considered without destroying the sample completely. On
the other hand densities of active particles could be higher in
other medium of interest. In the case of cw inputs in a cavity we
have an advantage that the medium noise is further reduced by a
factor of $\gamma/\Gamma$.

\section{Conclusion}

We developed a model based on Langevin equations which allowed us
to get a purely analytic description of the signal and the quantum
noise for Coherent  Anti-Stokes Raman Spectroscopy.  If we use
sequential pulses (as in FAST CARS) or we are close to two-photon
resonance the non-resonant terms would not be important and the
quantum noise arising from the finite lifetimes and coherence
times of the atoms/molecules would be a dominant source of noise.
When interaction time between the input fields and the medium is
short, we found that the signal and the medium noise have the same
behavior in free space or in a cavity, and with pulsed inputs or
with continuous waves.

In particular we showed that if the driving fields do not vary
much and the signal field is weak in comparison, the shot noise
which represents the fundamental limits of noise is larger than
the quantum noise due to the medium, and so the latter is not a
limiting factor. Using a cavity to achieve steady state with
continuous wave inputs leads to enhanced signal and lesser medium
noise; this would be important close to two-photon resonance and
the non-resonant terms do not contribute to the background noise.

Our calculations should be particularly relevant for novel
experiments with  newly developed femtosecond lasers and for fast
spectroscopic characterizations of microscopic agents in the air
which could be organic ones like anthrax spores or inorganic
suspensions or trace contaminants.

Although our calculations in this paper were specific to CARS, the
model we developed should be applicable to most coherent Raman
schemes with minor changes. We did not resort to the commonly used
adiabatic elimination when calculating the noise and we point out
the significant errors that would arise from such an
approximation. Thereby we set the grounds for a more accurate
understanding of quantum noise in stimulated Raman spectroscopy.
In the regime of short interaction times we achieved a completely
analytical description.  But the Langevin equations we set up in
our model describe a much broader physical regime, the solutions
in general will have to be found numerically.

\section*{Acknowledgments}

We thank Kevin Lehmann, Tomas Opatrny and Yuri Rostovtsev for
useful discussions and gratefully acknowledge the support from the
Office of Naval Research, the Air Force Research Laboratory (Rome,
NY), Defense Advanced Research Projects Agency-QuIST, Texas A$\&$M
University Telecommunication and Information Task Force (TITF)
Initiative, and the Robert A. Welch Foundation. One of us, Y. M.
Golubev acknowledges financial support from the following
organizations: RFBR (grant 03-02-16035), Minvuz of Russia (grant E
02-3.2-239), and by the Russian program  "Universities of Russia"
(grant ur.01.01.041).

\appendix

\section{Diffusion Coefficients}\label{app-diff}

A quantum Langevin equations for an operator $\hat{x}(t)$ has the
general structure
\bn \dot{\hat{x}}(t)=\hat{A}_x(t)+\hat{F}_x(t)\en
with a deterministic part $\hat{A}_x$ (not to be confused with the
field operator defined in Eq.~(\ref{defA})) and a stochastic part
$\hat{F}_x$.  The diffusion coefficients $2D_{xy}=\langle
F_xF_y\rangle$ associated with these equations are calculated
using the generalized fluctuation-dissipation theorem
\cite{Louisell,Lax}, often called the Einstein relations:
\bn\label{fdtheorem} d_t\langle\hat{x}\hat{y}\rangle=\langle
\hat{F}_{x}\hat{F}_{y}\rangle+ \langle\hat{x}\hat{A}_{y}\rangle
+\langle\hat{A}_{x}\hat{y}\rangle\en
The \emph{non-zero} diffusion coefficients, corresponding to
Eqs.~(\ref{microeqns}) for the microscopic atomic variables, are:
\bn  2D_{\sigma^\dagger \sigma}
&=&(2\Gamma-\Gamma_{b})\langle\hat{\sigma}_{b}\rangle
\n\\
2D_{\sigma\sigma^\dagger} &=&
2\Gamma\langle\hat{\sigma}_{c}\rangle+\Gamma_b
\langle\hat{\sigma}_{b}\rangle,
\n\\
 2D_{\sigma b}
=-2D_{\sigma c}&=&\Gamma_{b}\langle\hat{\sigma}\rangle \n\\
2D_{cc} =2D_{bb}=-2D_{bc}
&=&\Gamma_{b}\langle\hat{\sigma}_{b}\rangle.  \en
From these using the definitions in Eq.~(\ref{macrodef}) the
non-vanishing diffusion coefficients for the macroscopic
collective atomic operators $\hat{M}$ and
$\hat{N}=\hat{N}_b-\hat{N}_c$ are immediately obtained:
 \bn
2D_{NN}&=&4\Gamma_b\langle
\hat{N}_b\rangle,\n\\
2D_{\hat{M}^{\dagger}M}&=&(2\Gamma-\Gamma_b)\langle
\hat{N}_b\rangle, \n\\
2D_{MN}&=&2\Gamma_b\langle \hat{M}\rangle, \n\\
2D_{MM^{\dagger}}&=&2\Gamma \langle \hat{N}_c\rangle+\Gamma_b
\langle \hat{N}_b\rangle.\en
In transforming to c-numbers, by normal ordering, all the moments
for the noise must remain unaltered.  Gaussian noise is determined
by the first two moments; the first moment (i.e. the mean)
vanishes, the second moments must have the same time-evolution for
a pair of operators $\hat{x},\hat{y}$ in \emph{normal-order} and
their c-number equivalents:
\bn d_t\langle\hat{x}\hat{y}\rangle= d_t\langle xy\rangle.\en

The fluctuation-dissipation theorem Eq.~(\ref{fdtheorem}) and its
classical counterpart thereby relates the c-number diffusion
coefficients to the operator diffusion coefficients:
\bn \langle {\cal F}_{x}{\cal F}_{y}\rangle=\langle
\hat{F}_{x}\hat{F}_{y}\rangle\h{-0.5mm}+\h{-0.5mm}
\langle\hat{x}\hat{A}_{y}\rangle
\h{-0.5mm}+\h{-0.5mm}\langle\hat{A}_{x}\hat{y}\rangle\h{-0.5mm}-\h{-0.5mm}\langle
x{\cal A}_{y}\rangle \h{-0.5mm}-\h{-0.5mm}\langle{\cal
A}_{x}y\rangle.\en
Normal ordering of
$\langle\hat{x}\hat{A}_{y}\rangle+\langle\hat{A}_{x}\hat{y}\rangle$
gives some additional terms, and if $\hat{x},\hat{y}$ are
\emph{not} in normal order, we need to replace
$d_t\langle\hat{x}\hat{y}\rangle\rightarrow
d_t\langle\hat{y}\hat{x}\rangle=d_t\langle\hat{x}\hat{y}\rangle-d_t[x,y]$.
This determines all the non-vanishing c-number diffusion
coefficients for the collective atomic variables:
\bn\label{cnum_dc} 2{\cal D}_{{\cal N}{\cal N}}&=&2\Gamma_b\langle
N_T+{\cal
N}\rangle-4g\langle({\cal M}^*{\cal A}+{\cal A}^*{\cal M})\rangle,\n\\
2{\cal D}_{{\cal M}^*{\cal
M}}&=&(\Gamma-{\textstyle\frac{1}{2}}\Gamma_b)\langle N_T+{\cal
N}\rangle,\n\\ 2{\cal D}_{{\cal M}{\cal M}}&=&2g\langle {\cal
M}{\cal A}\rangle.\en

\section{Noise spectrum}\label{app-spectrum}

The Fourier transforms of the noise functions are defined by:
\bn F(\omega)=\frac{1}{2\pi}\int dt e^{i\omega t} F(t);\h{2mm}
F(t)=\int d\omega e^{-i\omega t} F(\omega). \en
For stationary processes the second moments or two-time
correlations depend only on the time difference: $\langle
F_i(t)F_j(t+\tau)\rangle=\Gamma(\tau)$. In the frequency domain
this defines the spectral density $P(\omega)$ of the noise:
\bn \langle F_i(\omega)F_j(\omega')\rangle
=\delta(\omega+\omega')P_{ij}(\omega)\en
The two-time correlation and the spectral densities are
time-frequency Fourier transforms of each other:
\bn P_{ij}(\omega)=\frac{1}{2\pi}\int d\tau
e^{i\omega\tau}\Gamma(\tau);\h{2mm} \Gamma(\tau)=\int d\tau
e^{-i\omega\tau}P_{ij}(\omega).\n\en
The fluctuations considered in this paper are characterized either
by $\delta$-correlations or by exponentially decaying correlations
as a function of time difference for which the spectral densities
are respectively constant and Lorentzian:
\bn
\Gamma(\tau)=2D_{ij}\delta(\tau)&\Rightarrow&P_{ij}=2D_{ij}\frac{1}{2\pi}\n\\
\Gamma(\tau)=e^{-\gamma|\tau|}&\Rightarrow&P_{ij}=\frac{2\gamma}{\gamma^2+\omega^2}\frac{1}{2\pi}.\en

\section{Propagating fields}\label{app-prop}

In order to describe fields propagating through a sample of atoms
in free space, we use the technique of Drummond and Carter
\cite{Drummond}.  We illustrate this technique by considering the
propagation of the signal field $\nu_4$, with wavenumber $k$ in
the direction of propagation (the z-axis); the part of the total
Hamiltonian involving this field is
\bn \hat{H}=\hbar\nu a_{4}^\dagger a_4+\hbar
g[\hat{a}_{4}\hat\sigma^{\dagger} +H.a.]\en
For simplicity we have assumed a real interaction strength
$g=R_{43}\alpha_3^{*}=|R_{43}\alpha_3^{*}|$. The interaction
length $L$ is divided into $2m+1$ equal segments of mean positions
$z_l= lL/(2m+1)$ for $l=-m,\cdots,m$.  Description of propagation
requires multiple modes; a natural basis is provided by the normal
modes for periodic boundary conditions on the quantization length
$L$
\bn k_n=\frac{2\pi n}{L}\h{1cm} n=-m,\cdots,m,\en
with corresponding creation and annihilation operators
$c_n^\dagger$ and $c_n$. Thus we can write the Hamiltonian as a
linear combination of the sub Hamiltonians for each discrete
segment summed over all the normal modes
\bn \hat{H}=\sum_n\hbar\nu c_n^\dagger c_n+\hbar
g\sum_{i,n,l}\left[c_n e^{-ik_nz_l}\hat\sigma^{\dagger(l)}
+H.a.\right]\en
The index $i$ accounts for the number of atoms in each segment,
$\sum_i=N_T/(2m+1)$. Introduce local operators for the slowly
varying field amplitude in each segment through a discrete Fourier
sum of the modes
\bn
\hat{a}_{4}^{(l)}=\frac{1}{\sqrt{2m+1}}\sum_{n=-m}^{m}c_ne^{ik_nz_l}\en
and the Hamiltonian can be written as
\bn \hat{H}=\sum_l\hbar\nu \hat{a}_{4}^{\dagger(l)}
\hat{a}_{4}^{(l)}+\sum_{ll'}\hbar\nu_{ll'}
\hat{a}_{4}^{\dagger(l)} \hat{a}_{4}^{(l')}\n\\+\hbar
g\sum_{l}\left[\sqrt{2m+1}\hat{a}_{4}^{\dagger(l)}
\hat\sigma^{\dagger(l)} +H.a.\right],\en
\bn \nu_{ll'}=\sum_{n=-m}^{m}\frac{2\pi n c}{(2m+1)L}\exp\left(
\frac{2\pi in(l-l')}{2m+1}\right).\en
The equation of motion for the slowly varying field amplitude is
then
\bn \dot{\hat{a}}_4^{(l)}=i\nu_{ll'}\hat{a}_4^{(l)}-ig
\sqrt{2m+1}\hat\sigma^{\dagger(i,l)}.\en
We convert to c-numbers and take the limit of $m\rightarrow\infty$
so that we have the following correspondence
\bn z_l=\frac{lL}{2m+1}&\rightarrow& z\n\\
\sqrt{2m+1}\alpha_4^{l}&\rightarrow& \alpha_4(z,t)\n\\
i\nu_{ll'}\sqrt{2m+1}&\rightarrow& c\frac{\partial}{\partial z}\n\\
-i\lim_{m\rightarrow\infty}(2m+1)\sum_i\sigma^{\dagger(l)}|_{z_l\rightarrow
z} &\rightarrow& {\cal M}(z,t)\en
and thereby we arrive at the space-time dependent equation of
motion for the slowly varying amplitude of the propagating field
\bn\label{spacetime} \left(\frac{\partial}{\partial
t}+\frac{c}{n}\frac{\partial}{\partial
z}\right)\alpha_4(z,t)=g{\cal M}(z,t).\en
In the same way the space-time dependent noise operators  are
defined by
\bn
\hat{F}_x(z,t)=\lim_{m\rightarrow\infty}(2m+1)\sum_i\hat{F}^{il}_x(t)\en
and analogously their c-number counterparts.  The space-time
dependent c-number diffusion correlations
\bn & &\langle {\cal F}_x(z,t){\cal
F}_y(z,t)\rangle\n\\
&=&\lim_{m\rightarrow\infty}(2m+1)^2\sum_{ij}\langle {\cal
F}^{il}_x(t){\cal F}^{jl'}_y(t)\rangle|_{z_l\rightarrow
z,z_l'\rightarrow z}\n\\
&=&L\langle {\cal F}_x(t){\cal F}_y(t)\rangle\delta(z-z'), \en
in the last step we used the correspondence
\bn
\lim_{m\rightarrow\infty}\frac{(2m+1)}{L}\delta_{ll'}=\delta(z-z').\en

\section{Problems in using adiabatic elimination in computing
noise}\label{app-adel}

We pointed out earlier that adiabatic elimination of atomic
variables cannot be applied to the fluctuations, we will now
briefly discuss the consequences of doing so.  We set
$1+B|\xi|^2\simeq 1$ as justified earlier.  We eliminate the
atomic variables and write the consequent equation for the field
variable inclusive of \emph{both} the mean and the fluctuation
\bn\label{eqA} \dot{\cal A}&=&-\gamma ({\cal A}-\xi)-C{\cal
A}+{\cal F_A} \\
{\cal F_A}&=& \frac{g}{\Gamma}\left[{\cal F}_{M}+\frac{g{\cal
A}}{\Gamma_b} \left\{{\cal F}_{N} -\frac{2g}{\Gamma}{\cal
F}_{T}\right\}\right]. \n\en
In the case of a cavity, using the $\delta(t-t')$ correlation of
the noise, we find
\bn  \delta{\cal S}^2(\Delta t_3)=g^2\frac{1-e^{-2\gamma \Delta
t_3}}{2\gamma}\left[\frac{2{\cal D_{TT}}}{\Gamma^2}+\frac{B|\xi|^4
2{\cal D_{NN}}}{\Gamma_b\Gamma} \right]\n\en
While this reproduces the long term steady-state behavior
correctly, the short-time behavior is linear in time, contrary to
the quadratic dependence that we found earlier.

The problem is more serious when we consider the free-space
problem.  We can understand it by considering the equation for the
fluctuation in the field variable
\bn (\partial_t +v\partial_z)\delta{\cal A}(z,t)&=&{\cal
F_A}(z,t).\en
On integrating in the frequency domain and using the boundary
condition $\delta{\cal A} (0,t)=0$ we find that
\bn \delta{\cal A}(L,t)&=&\frac{1}{v}\int_0^L dz {\cal
F_A}(z,t-\frac{n}{c}(L-z)).\en
As shown in  Appendix \ref{app-prop} the second moments of the
noise are delta-correlated in \emph{space} as well as in
\emph{time}
\bn\langle{\cal F_A}(z,t) {\cal F_A}(z',t')\rangle=2{\cal D_{AA}}
L\delta(z-z')\delta(t-t') \n\en
so that the second moments of the fluctuations of the signal at
the output are
\bn \langle\delta{\cal A}(L,t)\delta{\cal A}(L,t')\rangle =2{\cal
D_{AA}}\frac{L^2}{v^2}\delta(t-t').\en
At equal times this diverges, and therefore so does the variance
$\delta S^2$, and this underscores why the adiabatic approximation
is inappropriate for describing the fluctuations.

\section{Fluctuations for interaction in a cavity}\label{app-nonad}

We show the details of the derivation when the interaction takes
place in a cavity. From eqs.~(\ref{fluct-pc}) and (\ref{eqnT}) we
get
 \bn \delta{\cal S}^2(\Delta t) = g^2\int_0^{\Delta t}\h{-3mm}dt\int_0^{\Delta
t}\h{-3mm}dt'e^{-\gamma (2\Delta t -2t+\tau)}\h{2cm}\\
\times\frac{1}{2\pi}\int_{-\infty}^{\infty}\h{-3mm}d\Omega
e^{-i\Omega}\frac{\left[4g^4|\xi|^42{\cal D_{NN}}+g^2
 (\Gamma_b^2+\Omega^2)2{\cal D_{TT}}\right]}{(\Gamma^2+\Omega^2)(\Gamma_b^2+\Omega^2)
}\n\en
with $\tau=t-t'$. On doing a partial fraction decomposition and
carrying out the integrations
\bn  \delta{\cal S}^2(\Delta t) =\frac{2g^2}{2}\times\h{4.5cm} \n\\
\left[\frac{4g^2|\xi|^42{\cal D_{NN}}}{(\Gamma^2-\Gamma_b^2)}
\left(\frac{1-e^{-2\gamma \Delta
t}}{2\gamma(\gamma+\Gamma_b)\Gamma_b}- \frac{ e^{-2\gamma \Delta
t}-e^{-(\Gamma_b+\gamma)\Delta t}}{(\Gamma_b^2-\gamma^2)\Gamma_b}
\right.\right.\n\\\left.\left.- \frac{1-e^{-2\gamma \Delta
t}}{2\gamma(\gamma+\Gamma)\Gamma}+ \frac{ e^{-2\gamma \Delta
t}-e^{-(\Gamma+\gamma)\Delta
t}}{(\Gamma^2-\gamma^2)\Gamma_b}\right) \right.\n\\\left. +2{\cal
D_{TT}}\left(\frac{1-e^{-2\gamma \Delta
t}}{2\gamma(\gamma+\Gamma)\Gamma}+ \frac{ e^{-2\gamma \Delta
t}-e^{-(\Gamma+\gamma)\Delta
t}}{(\Gamma^2-\gamma^2)\Gamma_b}\right)\right].\ \ \en
In the limit of short interaction time $\gamma \Delta t\ll 1$, it
is easy to see that on doing a Taylor expansion of the
exponentials as a function of $\gamma \Delta t$ the zeroeth and
first order terms cancel out. For example consider the coefficient
of $2{\cal D_{TT}}$ in the last line which  expanded to second
order gives
\bn\frac{-2\gamma^2\Delta t ^2}{2\gamma(\gamma+\Gamma_b)\Gamma_b}-
\frac{4\gamma^2\Delta t^2-(\Gamma_b+\gamma)^2\Delta
t^2}{2(\Gamma_b-\gamma)(\gamma+\Gamma_b)\Gamma_b} = \frac{\Delta
t^2}{\Gamma}.\en
Thereby in this limit of short pulses we arrive at the expression
for $\delta S^2(\Delta t)$ in Eq.~(\ref{pulse-short}).

In the opposite limit of long time interaction times $\gamma
\Delta t\gg 1$ we get
\bn \delta{\cal S}^2(\Delta t) \simeq
\frac{g^2}{2}\left[\frac{2{\cal
D_{TT}}}{\gamma(\gamma+\Gamma)\Gamma} \h{3cm}\right.\n\\\left.+
\frac{4g^2|\xi|^42{\cal D_{NN}}}{(\Gamma^2-\Gamma_b^2)}
\left(\frac{1}{\gamma(\gamma+\Gamma_b)\Gamma_b}-
\frac{1}{\gamma(\gamma+\Gamma)\Gamma}\right) \right].\en
In the case of sequential pulsed input we allow for decay of the
coherence and the excitation and we get the expression in
Eq.~(\ref{pulse-long}).  In the case of continuous wave inputs if
we take take $\gamma\ll \Gamma,\Gamma_b$ we reproduce
Eq.~(\ref{cw-long}).

\end{document}